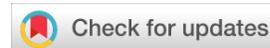

METHOD ARTICLE

REVISED

# A geospatial source selector for federated GeoSPARQL querying [version 2; peer review: 2 approved]

Antonis Troumpoukis [ID]1, Stasinos Konstantopoulos [ID]1, Nefeli Prokopaki-Kostopoulou [ID]2

1Institute of Informatics and Telecommunications, National Center for Scientific Research (NCSR) Demokritos, Ag. Paraskevi, 15341, Greece
2Department of Informatics and Telecommunications, National and Kapodistrian University of Athens, Athens, 16122, Greece

## Abstract

**Background**: Geospatial linked data brings into the scope of the Semantic Web and its technologies, a wealth of datasets that combine semantically-rich descriptions of resources with their geo-location. There are, however, various Semantic Web technologies where technical work is needed in order to achieve the full integration of geospatial data, and federated query processing is one of these technologies.

**Methods**: In this paper, we explore the idea of annotating data sources with a bounding polygon that summarizes the spatial extent of the resources in each data source, and of using such a summary as an (additional) source selection criterion in order to reduce the set of sources that will be tested as potentially holding relevant data. We present our source selection method, and we discuss its correctness and implementation.

**Results**: We evaluate the proposed source selection using three different types of summaries with different degrees of accuracy, against not using geospatial summaries. We use datasets and queries from a practical use case that combines crop-type data with water availability data for food security. The experimental results suggest that more complex summaries lead to slower source selection times, but also to more precise exclusion of unneeded sources. Moreover, we observe the source selection runtime is (partially or fully) recovered by shorter planning and execution runtimes. As a result, the federated sources are not burdened by pointless querying from the federation engine.

**Conclusions**: The evaluation draws on data and queries from the agroenvironmental domain and shows that our source selection method substantially improves the effectiveness of federated GeoSPARQL query processing.

## Open Peer Review

### Approval Status ✓ ✓

|  | 1 | 2 |
|---|---|---|
| **version 2** (revision) 06 Oct 2022 | ✓ view | ✓ view |
| **version 1** 19 Apr 2022 | ? view | ? view |

1. **Ozgu Can** [ID], Ege University, Bornova, Turkey

2. **Markus Stocker** [ID], TIB - Leibniz Information Centre for Science and Technology, Hannover, Germany

Any reports and responses or comments on the article can be found at the end of the article.

**Corresponding authors:** Antonis Troumpoukis (antru@iit.demokritos.gr), Stasinos Konstantopoulos (konstant@iit.demokritos.gr), Nefeli Prokopaki-Kostopoulou (nefelipk@di.uoa.gr)

**Author roles: Troumpoukis A**: Conceptualization, Data Curation, Investigation, Methodology, Software, Writing – Original Draft Preparation, Writing – Review & Editing; **Konstantopoulos S**: Conceptualization, Project Administration, Supervision, Writing – Review & Editing; **Prokopaki-Kostopoulou N**: Data Curation, Investigation, Resources, Writing – Original Draft Preparation

**Competing interests:** No competing interests were disclosed.

**Grant information:** This research was financially supported by the European Union's Horizon 2020 research and innovation programme under the grant agreement No 825258 (From Copernicus Big Data to Extreme Earth Analytics [ExtremeEarth]).
*The funders had no role in study design, data collection and analysis, decision to publish, or preparation of the manuscript.*

**How to cite this article:** Troumpoukis A, Konstantopoulos S and Prokopaki-Kostopoulou N. **A geospatial source selector for federated GeoSPARQL querying [version 2; peer review: 2 approved]** Open Research Europe 2022, **2**:48
https://doi.org/10.12688/openreseurope.14605.2

**First published:** 19 Apr 2022, **2**:48 https://doi.org/10.12688/openreseurope.14605.1

## Plain language summary

Nowadays, many data providers choose to publish their datasets as public services. This situation provides an opportunity to develop federated query processors (or federators), which are systems that combine several public sources as a single, virtual dataset. To serve an input query, a federator poses several queries on remote sources, and it combines their results accordingly.

The first problem a federator must solve to serve a query, is that of deciding which federated sources are relevant to which parts of the query. This problem is known as source selection, and the software that handles it as a source selector. A standard source selection technique is to annotate each source with information about the kind of data it contains, and use it to remove sources that have irrelevant data. For example, a source with snow data is irrelevant a query that requires fetching crop data.

Geospatial data is data about objects that have a location on the surface of the Earth. Not much work has been done in federating geospatial sources, and no source selection methods that target geospatial data currently exist.

We propose a new geospatial source selector for geospatial data. Since it is more likely for a source to contain data from a specific area (e.g., a country border) rather than the entire Earth, we can use such information for better source selection. For example, a geospatial source that contains fields within Austria is irrelevant for a query that requires crop data within an area in Greece.

We evaluate our method using data and queries from the agroenvironmental domain. Even though we spend more time in source selection, the number of sources used is smaller. This results in a more effective query processing, because the federator has to issue less queries to the remote endpoints to evaluate the query.

## Introduction

Geospatial data is primarily managed and processed by GIS systems that extend relational database management systems with geospatial relations. Although these systems are extremely efficient and optimized for the centralized processing of geospatial queries, they were not developed with the open and decentralized nature of Web data in mind. Semantic Web technologies, on the other hand, specifically facilitate the open, interoperable, and decentralized nature of Web data.

The Open Geospatial Consortium (OGC) is the global community that brings together GIS technology providers with geospatial data providers and consumers. OGC defines standards that improve access to geospatial data. In order to bring into the scope of geospatial databases the advantages offered by Semantic Web technologies, OGC specified the GeoSPARQL standard query language[1] for accessing databases that serve semantically-rich descriptions of resources and their geo-location. Besides being an OGC standard under active maintenance and update[2], GeoSPARQL has also seen implemented by database vendors in both the Semantic Web[3] and the RDBMS communities[4] demonstrating the geospatial community's increasing adoption of standards that facilitate the publication of interoperable data. There are, however, various Semantic Web technologies where technical work is needed in order to achieve the full integration of geospatial data[5].

Federated query processing is one such technology. Federated querying allows users to execute queries that combine information from several data sources, hiding the details of collecting and combing partial query results from each source. Source selection is the process of mapping each statement of the query to a subset of the SPARQL endpoints that make up a federation. Source selection is typically based on characteristic properties and URI namespaces to eliminate sources that do not have relevant data and dramatically improve the efficiency of federated query processing. This approach breaks down when geospatial datasets are distributed by geographical extent.

In this paper, we explore the spatial extent of each data source as a new type of summary. Spatial extent makes more sense for geospatial data, in comparison to the vocabularies and URI namespaces used which make more sense for thematic data. In practice, we investigate how to best exploit the fact that geospatial datasets are likely to be naturally divided in a canonical geographical grid[a] or following administrative regions or, more generally, areas of responsibility.[b] In the remainder of this paper, we first present a characteristic use case which we use throughout the paper (Section: Motivation and use case) and then provide background information (Section: Background). Next, we describe our geospatial source selector that uses endpoint metadata (in the form of a bounding box description) to flter out sources that do not contribute to the result (Section: The Source Selector). We then use our open-source implementation of this source selector to empirically compare the effciency of using bounding-box descriptions, precise and approximated shape descriptions, and conventional source selection (Section: Evaluation). Finally, we present and discuss relevant work (Section: Related work) and conclude (Section: Conclusions and future work).

## Contributions

- We present a new method for selecting data sources in a federated query processing context, specifcally targeting federations of geospatial databases. Our method is formally proven to be correct for arbitrary federations. That is, the method is guaranteed to not miss and relevant data source in any data segmentation, including overlapping datasets, multiple layers, and any such non-trivial situation.

- The method is confgured by an explicit formalism that summarizes the geospatial extent of each data source.

- We empirically compare three different ways to extract the summary of a geospatial dataset in this formalism: minimum bounding box, approximate shape, and exact shape. Our experiments demonstrate their relative strengths and weaknesses and guide decisions regarding which summary should be preferred depending on the characteristics of different summarized datasets.

- We compare geospatial extent-based data source selection to conventional thematic data source selection and demonstrate the signifcant improvement is selection accuracy that is achieved by using geospatial summaries to select data sources.

- We demonstrate that data source selection accuracy has signifcant impact on query processing time and that the overheads incurred by geospatial source selection are (at worst) compensated by reduced query execution times, with signifcant reduction of the overall execution time in some cases.

- We provide a prototype implementation of the proposed source selection method (including the summary extractors) integrated with the Semagrow federated query processor.

- We provide the all data, software, and complete experimental setup used in the experiments described here.

## Motivation and use case

We will now present a characteristic use case that both motivates some of our technical choices and backs our experimental setup with data and a query load: the food security use case of the ExtremeEarth project[c], within which the work presented here is conducted. In this use case, crop type information needs to be combined with nearby snowfall and snow storage, since irrigation largely depends on snow storage and seasonal release of fresh water.

The queries that are most relevant for this analysis are spatial within queries, spatial intersection queries, and within-distance queries: retrieving the land parcels with a given crop that are within, intersecting, or within a given maximum distance respectively from any snow-covered area, without requiring the exact distance. Notice that within-distance queries are considerably more computationally demanding than spatial overlap and inclusion queries that can be answered from the index. On the other hand, by comparison to queries that actually compute the distance, they offer themselves to aggressive optimization: Many instances can be discarded in advance as they are too far away to be within the required distance so that the (expensive) distance computations actually performed by the database are minimized.

---

[a]Consider how weather and climate datasets are organized in a canonical grid, such as for example the Copernicus CMIP6 climate projections.

[b]Consider, for example, the offcial data portal of the EU or the UCSD GIS data portal that both organize datasets following administrative regions.

Consider, for example, Figure 1. Evaluating a filter that only retains shapes within distance $d$ from $p$ can immediately (*i.e.,* from the database index) discard all shapes contained in $s_1$ if the distance between $s_1$ and $p$ is greater than $d$. This presents a huge optimization opportunity by comparison to computing the distance between $p$ and all shapes in $s_1$ and then comparing these against $d$. Transferring this discussion to federated query processing, we see that geospatial datasets are often published by public administrations or other entities with responsibility over a specific geographic extent. This motivates applying this optimization to the source selection level: if $s_1$ and $s_2$ were the bounding polygons of all resources served by two GeoSPARQL endpoints, then source selection can exclude $s_1$ from the execution plan.

Naturally, a GeoSPARQL query will normally combine geospatial restrictions with thematic triple patterns; in our case, for example, referring to a crops code list or hierarchy. Consider, now, Figure 2 where each of the three data sources only contains triples using a specific code list or vocabulary. Such a situation is likely to appear when different organizations publish data regarding different aspects of a geographical region (for example, crops and precipitation data), some of which are also independently published for each region. For a query using only the 'green' vocabulary to retrieve entities of interest within a given distance from $p$, it makes no sense to consider including $s_1$ in the execution plan. This partitioning is amenable to conventional source selection based on metadata about the vocabularies used in each data source. However, in order to have the optimal source selection a federation engine would need to also exclude $s_3$ based on its geospatial extent, and only pose a query in $s_2$. Such a source selection can only be achieved by extending conventional federated source selection with a mechanism that combines metadata about the thematic content of a data source with metadata about its geospatial extent.

The queries discussed in the previous paragraphs are used for fetching data that have a specific spatial relationship with a fixed polygon $p$ in the query. However, a source selection mechanism that is aware of the geospatial nature of the sources, can be helpful and in queries that involve geospatial joins. Consider, for example, the federation that consists of the four geospatial sources of Figure 3, and assume that we are interested in finding pairs of 'red' and 'blue' entities that their distance is less than $d$. As previously, we should exclude $s_4$ from the query since this source contains only 'green' entities. Suppose now that the distance between the boundaries of $s_1$ and $s_3$ is greater than $d$. Then, since all 'red' shapes are found in $s_1$, and the distance between $s_1$ and $s_3$ is greater than $d$, we can safely deduce that there does not exist any 'blue' shape in $s_3$ that is located within distance $d$ for all 'red' shapes. As a result, it makes sense from a practical point of view to exclude $s_3$ from the evaluation of the query because it contains irrelevant shapes; and to query only $s_1$ and $s_2$.

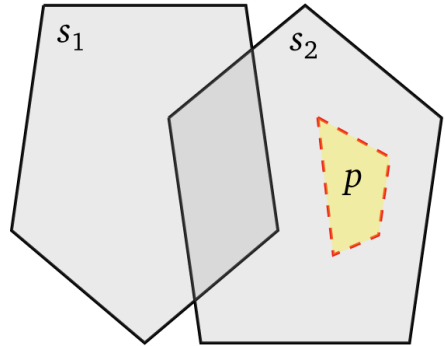

**Figure 1. The boundaries of two sources $s_1$ and $s_2$, with a polygon of interest $p$ that lies within the boundary of $s_2$.**

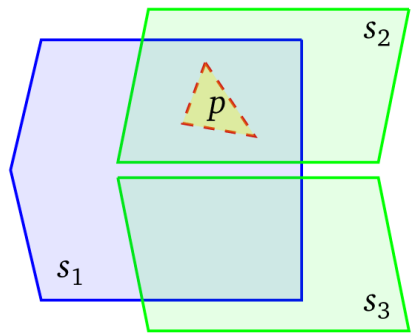

**Figure 2. The boundaries of three sources and a polygon of interest; $s_1$ uses the 'blue' vocabulary; $s_2$ and $s_3$ use the 'green' vocabulary.**

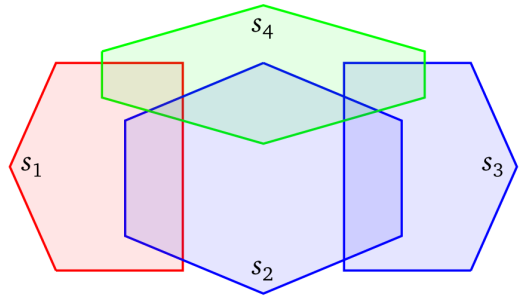

**Figure 3.** Four sources, where $s_1$ uses the 'red' vocabulary to describe resources, $s_2$ and $s_3$ the 'blue' vocabulary, and $s_4$ the 'green' vocabulary.

## Background

In this section, we provide the background of our approach. In particular, we present a brief introduction on the GeoSPARQL query language, which is an extension of SPARQL and the de-facto query language for querying Geospatial Linked Data[1]; and then we summarize the state-of-the-art on source selection in federated SPARQL query processors. Throughout the paper, we use SPARQL qnames to shorten URIs. The list of URI namespaces that we use are shown in Table 1.

### The GeoSPARQL query language

The GeoSPARQL specification[1] defines a set of classes and properties for asserting and querying geospatial information. The `geo:SpatialObject` class comprises any resource that can have a spatial representation. All the usual topological relations (containment, overlap, etc.) are foreseen as properties.

The `geo:SpatialObject` class subsumes the `geo:Feature` class that represents geo-located things that exist in the physical world; and `geo:Geometry` that represents spatial objects that have a single, concrete geographical shape. Each feature is linked with one (or more, *e.g.*, seasonal variation) geometries with the `geo:hasGeometry` property. The `geo:asWKT` property is used to provide the concrete geographical shape of any spatial object as an RDF literal of the `geo:wktLiteral` datatype.[c] Such RDF literals are usually referred as WKT literals, because they represent geometries in the WKT (*i.e.*, well-known text) format. Given the above, the link between features and their concrete coordinates follows the pattern:

```
r geo:hasGeometry g .
g geo:asWKT "WKT"^^geo:wktLiteral .
```

where `r` is an instance of `geo:Feature` and `g` is an instance of `geo:Geometry`.[d]

Naturally, inference about `geo:wktLiteral` values falls outside RDF graph entailment and can only be performed by specialized geospatial databases. Such entailment is accessed via geospatial functions. For example:

```
SELECT ?r WHERE {
  ?r geo:hasGeometry ?g .
  ?g geo:asWKT ?w .
  FILTER( geof:sfWithin(?w, "<http://www.opengis.net/def/crs/EPSG/0/4326>
      POLYGON((19.2477876 34.7006096,19.2477876 41.7488862,29.7296986 41.7488862,
      29.7296986 34.7006096,19.2477876 34.7006096))"^^geo:wktLiteral) )
}
```

---

[c] Two alternative serializations are foreseen by GeoSPARQL, `geo:wktLiteral` and `geo:gmlLiteral`, and two datatype properties, `geo: hasWKT` and `geo:hasGML`. We restrict the discussion in this paper to the WKT serialization, and it is straightforward to transfer this discussion to GML or any other serialization.

[d] Temporal or other restrictions might apply to select the correct `geo:Geometry` instance. We gloss over such considerations that fall well outside the scope of source selection based on geospatial extent.

This uses the `geof:sfWithin` function to access the geospatial operator that computes if the WKT value `?w` retrieved from the graph pattern is contained in another WKT value; such a query fetches all features within a given polygon.

## Source selection in federated querying

Federated query processors are systems that seamlessly integrate data from multiple remote dataset servers. A federated query processor receives a query, issues the necessary subqueries in the remote endpoints, combines the intermediate results accordingly, and presents the result to the client. Usually, federated query processing consists of three major phases, namely source selection, query planning, and query execution, as shown in Figure 4. In particular, source selection identifies which of the federated sources refer to which parts of the input query. Then, the query planner uses the output of the source selector in order to construct an efficient query execution plan which contains a set of subqueries to be issued in the source endpoints and how the inter- mediate results of the subqueries are combined. Finally, the query executor evaluates this query execution plan and constructs the query result to be sent as a response to the client.

The first step in federated SPARQL query processing is to select a subset of the sources that make up a federa- tion for each triple pattern of the query. The goal for the source selector is to prune as many irrelevant sources as possible in order for the query planner to come up with a more efficient query execution plan.

Most federated SPARQL query processors make use of two basic approaches for their source selection mechanism. In metadata-assisted source selection, the federator relies on a dataset descriptor about properties and classes for each federated source (for instance, expressed using the Vocabulary of Interlinked Datasets (VoID)[7]) in order to identify candidate sources for each individual triple pattern of the query[8]. On the other hand, in metadata-free source selection, first introduced by FedX[9], the federator identifies the candidate sources by issu- ing an ASK SPARQL query to all federated endpoints for each triple pattern of the query (and also uses a cache so that when the same triple pattern reoccurs, it doesn't have to ping the sources again). In general, the lat- ter approach is more accurate (because it relies on ASK queries and not metadata) but the former approach is faster (because issuing an ASK query to endpoints introduces a significant time overhead when a triple pat- tern appears for the first time and the cache cannot be used). To get the best of both worlds, most federation engines[10–14] use source metadata to perform a first pruning of the sources and then refine it using ASK queries.

Subsequent developments in source selection methodology for federated SPARQL systems provided new join-aware techniques that consider the input query as a whole[15–18] rather than isolated triple patterns. In this family of approaches, sophisticated source metadata about subject and object URI namespaces[19], or about characteristic sets that can describe complete star patterns[20] are able to support inferences on join variables to eliminate irrelevant sources from the plan.

**Table 1. List of URI namespaces used throughout the paper.**

| Prefix | Namespace |
|--------|-----------|
| `rdf:` | `<http://www.w3.org/1999/02/22-rdf-syntax-ns#>` |
| `rdfs:` | `<http://www.w3.org/2000/01/rdf-schema#>` |
| `geo:` | `<http://www.opengis.net/ont/geosparql#>` |
| `geof:` | `<http://www.opengis.net/def/function/geosparql/>` |
| `uom:` | `<http://www.opengis.net/def/uom/OGC/1.0/>` |
| `void:` | `<http://rdfs.org/ns/void#>` |
| `svd:` | `<http://www.w3.org/2015/03/sevod#>` |

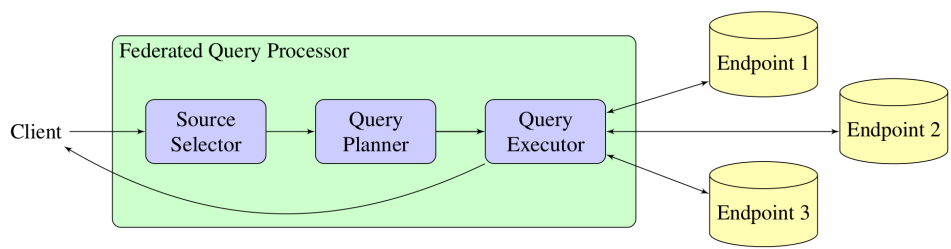

**Figure 4. Typical architecture of a federated query processor.**

We will close this section with some examples regarding the source selectors discussed above. A VoID-metadata-assisted source selector can decide whether to keep a source $d$ for the pattern `?s  p  ?o` by simply checking if `p` appears in the metadata of $d$. On the other hand, for the pattern `?s  ?p  "str"`, an ASK-based source selector may be more accurate, because the VoID vodabulary does not allow a user to define specific metadata for every literal that appears in the sources. Finally, consider the join `?s p1 ?x . ?s p2 ?y` and a candidate source $d$ for the first pattern. If the subject namespace of all triples of $d$ that match the first pattern is not included in the subject namespaces of all triples that match the second one in all of its candidate sources, then $d$ is a irrelevant source for the first pattern.

## The source selector

In this section, we present in detail our method, the source metadata it operates upon, and the method's implementation.

### Preliminaries

Let $a$ and $b$ be two spatial objects, that can be points, lines and/or polygonal areas. We say that $a$ and $b$ are *disjoint* if they do not have any points in common. Moreover, we say that $a$ contains $b$ if no points of $b$ lie in the exterior of $a$, and at least one point of the interior of $b$ lies in the interior of $a$. Notice that in this case, $a$ does not contain its border, but $a$ does contain itself. We say that $a$ and $b$ touch if they have at least one boundary point in common, but they share no interior points.

Let $Q$ be a GeoSPARQL query and $F$ be the set of sources federated by a query processor. We say that $S \subseteq F$ is the optimal source set for $Q$ if (a) using only the sources in $S$ gives the same results for $Q$ as using all the sources in $F$; and (b) there is no subset of $S$ that gives the same results for $Q$ as using all the sources in $F$.

### Source metadata

The proposed source selector requires every federated source to be tagged with the following information:

**Definition 1.** A bounding polygon $\mathscr{B}(s)$ of a geospatial source $s$ is any polygon that contains every spatial object in $s$.

For representing the bounding polygon of a source $s$, we extend the Sevod vocabulary[21] by defining the following property:[e]

**Definition 2.** The geometry of a bounding polygon of a dataset can be denoted using the predicate `svd:boundingWKT` as follows:

```
svd:boundingWKT    rdf:type       rdf:Property   ;
                   rdfs:domain    void:Dataset   ;
                   rdfs:range     geo:wktLiteral .
```

This property allows `void:Dataset` objects to be annotated with a concrete shape. As will become obvious below, our method relies on concrete bounding polygons for which distances can be computed, as opposed to geometry objects that might only be defined via geospatial relations.

An example of such an annotation is shown below.

**Example 1.** The following set of RDF triples represents a source which is accessible from a specific endpoint and contains only shapes within a specific polygon, denoted as a WKT literal:

```
[] rdf:type              void:Dataset ;
   void:sparqlEndpoint   <http://example.org/sparql> ;
   svd:boundingWKT       "<http://www.opengis.net/def/crs/EPSG/0/4326>
     POLYGON ((9.53155824986118  46.4017516462893, 9.53155824986118 49.0185728029906,
     17.1618132052086 49.0185728029906,  17.1618132052086 46.4017516462893,
     9.53155824986118 46.4017516462893)) "^^geo:wktLiteral .
```

---

Obtaining a bounding polygon of a geospatial RDF dataset is a simple task (even though obtaining a good one isn't). The most straightforward solution is to calculate the spatial union of all shapes in the dataset (*i.e.*, all objects of the `geo:asWKT` property), which produces an accurate bounding polygon of the geospatial source. The main drawback of this approach is that the resulting summary may be large in size (since it comprises a large number of coordinates). Alternatively, we can use less accurate bounding polygons that are expected to have smaller size. Examples include the minimum bounding box[22] of the spatial union, or an approximation of the spatial union using a quadtree[23] of height *k* (*i.e.*, a polygon obtained by partitioning the minimum bounding box of the spatial union in $4^k$ equal rectangles, by removing all rectangles that are disjoint from the spatial union and then calculating the union of the remaining rectangles). This trade-off between summary size and summary accuracy is one of the issues our paper discusses and will be revisited in the evaluation section.

### Source selection algorithm

In this subsection, we describe our source selection algorithm in detail. We assume the existence of some helper routines that their implementations is not included as a separate algorithm. Given a GeoSPARQL query *Q*, the routine GEOSPATIALFILTERS(*Q*) returns the set of all GeoSPARQL filters that appear in *Q*. Given a filter *f*, the routine VARS(*f*) returns all variables that appear in *f*. Given a filter condition without free variables *c*, EVAL(*c*) returns true if the condition holds, otherwise `false`. Finally, our algorithm builds on top of a thematic source selector, therefore we assume its existence by using the routine THEMATICSOURCESELECTOR (cf. Background section for a discussion regarding the state-of-the-art of thematic source selection in federated query processing).

Our source selector is designed to support a family of GeoSPARQL queries. We concentrate on simple filter expressions that consist of a geospatial function call. In particular, we consider filters whose condition is of the form *r*(*x*, *y*) or `geof:distance`(*x*, *y*, *u*) op *d*, where *x*, *y* are variables or WKT literals, *r* is a non-disjoint geospatial function (i.e., one of the `geof:` functions `sfEquals`, `sfContains`, `sfIntersects`, `sfOverlaps`, `sfCrosses`, or `sfWithin`), *u* is unit of measure, op $\in \{<, \leq, =\}$, and *d* is a numeric literal. Our method can be easily extended for the other GeoSPARQL functions apart from the simple features relation family (*e.g.*, Egenhofer[24] and RCC8[25]).

Algorithm 1 takes as input a GeoSPARQL filter and a boundary for each free variable of the geospatial filter, and returns true if the condition of the filter does not hold if we substitute every free variable of the filter with every shape contained in the corresponding boundary. We call this algorithm BPFILTEREMPTY, because it can check whether a GeoSPARQL filter *f* will return an empty result set, provided that all bindings of every variable of the filter are contained within a known polygon.

Algorithm 2 prunes the set of sources for every triple pattern of the form (*x*, `geo:asWKT`, *o*), by using the relevant GeoSPARQL filters of the triple pattern and the bounding polygons of each candidate source. Notice

---

**Algorithm 1. BPFILTEREMPTY**

**Input:**
    a GeoSPARQL filter *f* s.t. VARS(*f*) = $\{v_1, \ldots, v_n\}$,
    and a set of bindings $B = \{v_1/b_1, \ldots, v_n/b_n\}$.

**Output:** true or false

1: $R_1$ := {geof:sfEquals, geof:sfWithin, geof:sfContains}
2: $R_2$ := {geof:sfOverlaps, geof:sfCrosses, geof:sfTouches, geof:sfIntersects}
3: *f'* is a GeoSPARQL filter obtained by *f* by substituting $v_1, \ldots, v_n$ with $b_1, \ldots, b_n$.
4: **if** *f'* = *r*(*x*, *y*), where $r \in R_1$ **then**
5:     **return** EVAL(geof:sfDisjoint(*x*, *y*)) $\lor$ EVAL(geof: sfTouches(*x*, *y*))
6: **else if** *f'* = *r*(*x*, *y*), where $r \in R_2$ **then**
7:     **return** EVAL(geof:sfDisjoint(*x*, *y*))
8: **else if** *f'* = geof:distance(*x*, *y*, *u*) op *d*,
        where *u* is unit of measure, op $\in \{<, \leq, =\}$, and *d* is a numeric literal **then**
9:     **return** EVAL(geof:distance(*x*, *y*, *u*) > *d*)
10: **else**
11:     **return** false
12: **end if**

---

**Algorithm 2. AsWktSourceSelector**

**Input:**
    a GeoSPARQL query $Q$,
    a set $T$ of triple patterns,
    a set $S$ of sources,
    a mapping $\sigma : T \rightarrow 2^S$,
    a mapping $\mathscr{B} : S \rightarrow B$ s.t. $\mathscr{B}(s)$ is the bounding polygon of $s$

**Output:** a mapping $\sigma : T \rightarrow 2^S$

1:  $F \coloneqq \text{GeospatialFilters}(Q)$
2: **repeat**
3:     $\sigma_{old} \coloneqq \sigma$
4:     **if** there exists some $f \in F$, $t \in T$, and $s \in S$ such that
           $\text{Vars}(f) = \{o\}$, $t = (x, \text{geo:asWKT}, o)$, $s \in \sigma(t)$
           and $\text{BPFilterEmpty}(f, \{o/\mathscr{B}(s)\})$ **then**
5:         $\sigma(t) \coloneqq \sigma(t) - \{s\}$
6:     **end if**
7:     **if** there exists some $f \in F$, $t, t' \in T$, and $s \in S$ such that
           $\text{Vars}(f) = \{o, o'\}$, $t = (x, \text{geo:asWKT}, o)$, $t' = (x', \text{geo:asWKT}, o')$, $s \in \sigma(t)$,
           and for all $s' \in \sigma(t')$ it holds $\text{BPFilterEmpty}(f, \{o/\mathscr{B}(s), o'/\mathscr{B}(s')\})$ **then**
8:         $\sigma(t) \coloneqq \sigma(t) - \{s\}$
9:     **end if**
10: **until** $\sigma_{old} = \sigma$
11: **return** $\sigma$

---

that AsWktSourceSelector uses BPFilterEmpty as a helper algorithm to check if a candidate source for a triple pattern would contain irrelevant data and thus it can be pruned. In the following, we show how AsWktSourceSelector works with a simple example:

**Example 2.** Let $s_1$, $s_2$, $s_3$ be three sources, $b_1$, $b_2$, $b_3$ be their bounding polygons respectively, and POLY be a WKT literal. Moreover, assume that $b_3$ is disjoint from both $b_1$ and $b_2$, and that POLY is within $b_1$ and $b_2$. Notice that POLY is disjoint from $b_3$. Now, consider the following GeoSPARQL query:

```
SELECT * WHERE {
  ?u geo:asWKT ?x .
  ?v geo:asWKT ?y .
  FILTER( geof:sfWithin(?x, POLY) )
  FILTER( geof:sfIntersects(?x, ?y) )
}
```

POLY is disjoint from $b_3$ and $b_3$ is the bounding polygon of $s_3$. Thus, no shape in $s_3$ can be within POLY. In other words, for all bindings of `?x` that come from $s_3$, the condition of first geospatial filter should return `false`. Therefore, $s_3$ contains irrelevant bindings for `?x`, and as a result $s_3$ can be pruned from the candidate sources of the first triple pattern. Indeed, it is easy to check that it holds:

$$\text{BPFilterEmpty}(\texttt{geof:sfWithin(?x, POLY)}, \{\texttt{?x}/b_3\}) = \text{true}$$

Source $s_3$ can be pruned from the sources of the second triple pattern as well. To check this, apart from the bindings of `?y` (as previously), we should consider the bindings of `?x` as well, which come from the candidate sources of the first triple pattern. $b_3$ is disjoint from both $b_1$ and $b_2$. Thus, no shape that belongs in $s_3$ can intersect with some shape in $s_1 \cup s_2$. In other words, for all bindings of `?y` that come from $s_3$, the condition of the second geospatial filter should return `false`, because all bindings of `?x` come from solely $s_1$ and $s_2$. Indeed, it is easy to check that it holds using:

$$\text{BPF{\scriptsize ILTER}E{\scriptsize MPTY}}(\texttt{geof:sfIntersects(?x, ?y)}, \ \{?x/b^*, ?y/b_3\}) = \texttt{true}, \text{ for all } b^* \in \{b_1, b_2\}$$

As a result, A{\scriptsize S}W{\scriptsize KT}S{\scriptsize OURCE}S{\scriptsize ELECTOR} should prune $s_3$ from the candidate sources of both triple patterns of the query.

Algorithm 3 makes use of the previous algorithms and defines the proposed source selection mechanism. Notice that A{\scriptsize S}W{\scriptsize KT}S{\scriptsize OURCE}S{\scriptsize ELECTOR} targets only triple patterns that link a specific geometry URI with its WKT serialization, *i.e.*, patterns of predicate `geo:asWKT`. It leaves out triple patterns that link a resource with its geometry (*i.e.*, `geo:hasGeometry` predicates), and triple patterns of thematic information. As discussed previously, the geospatial pruning obtained by Algorithm 2 should be complemented by a thematic source selector. Here we use the URI-prefix-based source selector included in Semagrow (which we will discuss in the implementation subsection). As a final note, the result of the algorithm is a mapping that associates each triple pattern of the query to a subset of the set of the sources of the federation. Since, in practice, source selectors are built 'on top' of each other, our source selector takes such a mapping as input, which can be the output of another source selector.

## Correctness of source selection
In the following, we discuss the correctness of our source selection. In particular, we show that the geospatial source selector does not remove any source that contains necessary data for the evaluation of the query, *i.e.*, it does not remove any source that belongs to the optimal source set of the query. We begin with an important property of Algorithm 1, and then we show that the set of sources that are kept in the output of the source selectors of Algorithm 2 and Algorithm 3

**Lemma 1.** *Let f be a GeoSPARQL filter such that* V{\scriptsize ARS}$(f) = \{v_1, \ldots, v_n\}$, *and* $B = \{v_1/b_1, \ldots, v_n/b_n\}$ *be a set of variable bindings for all free variables of f. Then,* BPF{\scriptsize ILTER}E{\scriptsize MPTY}$(f, B) = \texttt{true}$ *if the condition of f does not hold if we substitute $v_i$ in f with any possible shape contained in $b_i$, for all $1 \le i \le n$.*

*Proof.* Notice that if two shapes *x*, *y* are disjoint, then every shape contained in *x* will also be disjoint from every shape contained in *y* (thus every pair of shapes taken from *x* and *y* are not related with any non-disjoint spatial relation). Moreover, if *x* touches *y*, then there does not exist any shape contained in *x* that is equal, within, or contains any shape contained in *y* and vice-versa. Finally, if the distance between *x* and *y* is greater than *D*, then the distance between every shape contained in *x* from every shape contained in *y* is also greater than *D*.

**Lemma 2.** *Let Q be a GeoSPARQL query, T be the set of triple patterns that appear in Q, S be a set of sources, $\mathcal{B}$ be a mapping such that $\mathcal{B}(s)$ is the bounding polygon of the source s, $\sigma$ be a mapping that maps each triple pattern of Q to a set of sources from Q. Then,* A{\scriptsize S}W{\scriptsize KT}S{\scriptsize OURCE}S{\scriptsize ELECTOR}$(Q, T, S, \sigma, \mathcal{B})$ *does not prune any source that belongs to the optimal source set of Q.*

*Proof.* Let *s* be a source pruned by the algorithm. It suffices to show that *s* does not belong in the optimal source set of *Q*. We distinguish the following cases:

---

**Algorithm 3.** G{\scriptsize EOSPATIAL}S{\scriptsize OURCE}S{\scriptsize ELECTOR}

**Input:**
    a GeoSPARQL query *Q*,
    a set *T* of triple patterns,
    a set *S* of sources,
    a mapping $\sigma : T \to 2^S$,
    a mapping $\mathcal{B} : S \to B$ s.t. $\mathcal{B}(s)$ is the bounding polygon of *s*,
    and thematic metadata $\mathscr{P}$
**Output:** a mapping $\sigma : T \to 2^S$
1:  $\sigma := $ A{\scriptsize S}W{\scriptsize KT}S{\scriptsize OURCE}S{\scriptsize ELECTOR}$(Q, T, S, \sigma, \mathcal{B})$
2:  $\sigma := $ T{\scriptsize HEMATIC}S{\scriptsize OURCE}S{\scriptsize ELECTOR}$(T, S, \sigma, \mathscr{P})$
3:  **return** $\sigma$

---

*Case 1:* There exists a geospatial filter $f$ of $Q$ such that VARS($f$) = {$o$} and a triple pattern $t = (x,$ geo:asWKT, $o$) of $Q$ such that $s \in \sigma(t)$ and BPFILTEREMPTY($f$, {$o/\mathscr{B}(s)$}) = true. According to Lemma 1, there does not exist any shape $z \in \mathscr{B}(s)$ s.t. the condition of $f$ holds. Since $s$ is the bounding polygon $\mathscr{B}(s)$ (and thus all shapes of $s$ are contained in $\mathscr{B}(s)$), it is easy to see that the result set of $Q$ in this case will not contain any bindings for $o$ that come from $s$; therefore, $s$ is not contained in the optimal source set of $t$.

*Case 2:* There exists a geospatial filter $f$ of $Q$ such that VARS($f$) = {$o$, $o'$} and two triple patterns $t = (x,$ geo:asWKT, $o$), $t' = (x',$ geo:asWKT, $o'$) of $Q$, such that $s \in \sigma(t)$ and for all $s' \in \sigma(t')$ it holds BPFILTEREMPTY($f$, {$o/\mathscr{B}(s)$, $o'/\mathscr{B}(s')$}) = true. We denote by $U$ the *union* of the bounding polygons for all candidate sources of $t'$. According to Lemma 1, there does not exist any pairs of shapes $z \in \mathscr{B}(s)$ and $z' \in U$ s.t. the condition of $f$ holds. It is therefore clear that the result set of $Q$ will not contain any bindings for $o$ that come from $s$, because they are irrelevant to any binding for $o'$ that come from all sources of $t'$; therefore, $s$ is not contained in the optimal source set of $t$.

All thematic source selectors already proposed in the literature are correct (*i.e.,* do not prune sources that belong to the optimal source set of a query), thus:

**Theorem 1.** *Let $Q$ be a GeoSPARQL query, $T$ be the set of triple patterns that appear in $Q$, $S$ be a set of sources, $\mathscr{B}$ be a mapping such that $\mathscr{B}(s)$ is the bounding polygon of the source $s$, $\mathscr{P}$ are thematic metadata, $\sigma$ be a mapping that maps each triple pattern of $Q$ to a set of sources from $Q$. Then,* GEOSPATIALSOURCESELECTOR($Q$, $T$, $S$, $\sigma$, $\mathscr{B}$, $\mathscr{P}$) *does not prune any source that belongs to the optimal source set of $Q$.*

*Proof.* Using Lemma 2, and the fact that THEMATICSOURCESELECTOR does not prune any source that belongs to the optimal source set of $Q$.

### Illustrative Example

In this subsection, we use an example to illustrate how the geospatial source selector operates in practice. We use a federation of three geospatial sources: s1 contains "blue" shapes (i.e., resources of type ex:blue and their geometries); s2 and s3 contain "green" shapes (i.e., resources of type ex:green and their geometries). The boundary of s1 intersects with both the boundaries of s2 and s3, but s2 and s3 are disjoint (cf. Figure 2).

Consider the following query, that fetches all green shapes that intersect a given polygon of interest p which lies within the bounding polygon of both s1 and s2:

```
SELECT * WHERE {
  ?s rdf:type ex:green .      # t1
  ?s geo:hasGeometry ?g .     # t2
  ?g geo:asWKT ?w .           # t3
  FILTER( geof:sfIntersects(?w, "WKT of P"^^geo:wktLiteral) )
}
```

Intuitively, the optimal source set for all triple patterns should contain only $s_2$, because $s_1$ contains blue shapes and p lies outsides the bounding polygon of $s_3$.

In the following, we will show how GEOSPATIALSOURCESELECTOR operates in practice. We initialize all triple patterns with all the data sources in the federation:

$$\{(t_1, \{s_1, s_2, s_3\}), (t_2, \{s_1, s_2, s_3\}), (t_3, \{s_1, s_2, s_3\})\}$$

Then ASWKTSOURCESELECTOR is called. Since the polygon of interest p and the bounding polygon $\mathscr{B}(s_3)$ of $s_3$ are disjoint, Algorithm 2 prunes $s_3$ from $t_3$ as follows:

$$\{(t_1, \{s_1, s_2, s_3\}), (t_2, \{s_1, s_2, s_3\}), (t_3, \{s_1, s_2\})\}$$

Then, THEMATICSOURCESELECTOR is called. At first, the thematic source selection identifies (e.g., using predicate and class metadata) that $t_1$ contains no bindings in $s_1$ (notice that no triples with object ex:blue appear in $s_1$):

$$\{(t_1, \{s_2, s_3\}), (t_2, \{s_1, s_2, s_3\}), (t_3, \{s_1, s_2\})\}$$

In the second phase, we apply the Hibiscus join-aware thematic source selector (cf. Background section), specifically configured to exploit the assumption that `geo:hasGeometry - geo:asWKT` paths are served from the same data source as the resource they annotate. Under this assumption, in our query, $t_1$, $t_2$ and $t_3$ must be served from the same set of data sources:

$$\{(t_1, \{s_2\}), (t_2, \{s_2\}), (t_3, \{s_2\})\}$$

We observe that we reached the intuitive result mentioned previously, that is, $s_2$ is the only source assigned to all triple patterns of the query.

### Implementation

We provide an implementation of our geospatial source selector integrated in the Semagrow SPARQL federation engine. The elementary geospatial operations used in our source selector implementation are provided by the rdf4j framework. Our selector wraps the behaviour of the purely thematic HiBISCuS source selection mechanism[19], used by Semagrow version 2.1.0 or newer (we note that the original version of Semagrow[11] did not use it). As a result, the underlying thematic source selector of Semagrow uses both triple-pattern-wise source selection (using predicate and class metadata and ASK queries) and join-aware source selection (using URI-prefix subject and object metadata).

Moreover, we provide a tool (sevod-scraper) for extracting the source metadata required by our source selector. It takes, as input, an RDF dump and calculates a bounding polygon of the dataset, which can be a) the spatial union of all shapes in the dataset, b) the minimum bounding box of all shapes in the dataset, or c) an approximation of the union of all shapes in the dataset using a quadtree of given height.

## Evaluation

In this section, we evaluate the performance of the proposed source selector. We describe in detail the experimental setup and we analyze the results. Our experiment is based on real-world datasets and is inspired by a practical use-case scenario in the domain of food security.

### Experimental setup

In the following we describe the experimental setup of our evaluation, *i.e.,* the data used, the source endpoints, the configuration of the federations that are compared, the queries of the experiment, and some details on the experiment deployment and execution.

**Datasets** For the experimental evaluation, we use the following data sources:

1. The Database of Global Administrative Areas (GADM) for Austria, which contains all administrative divisions of Austria up to Level-3.

2. The Austrian Land Parcel Identification System (INVEKOS), which contains the geo-locations of all crop parcels in Austria and the owners' self-declaration about the crops grown in each parcel.

3. A snow cover map, which contains thematic and geospatial snow data within Austria from February to April of 2018.

We envisage that Austrian state governments publish crop data for their own area of responsibility; and a further (possibly different) entity publishes snow cover datasets for the same area. As the datasets described previously refer to the whole region of Austria, we partitioned them for the purposes of our experiment to datasets that refer to smaller areas. Regarding the administrative and crop datasets, we partition them into smaller datasets according to the polygons of the states of Austria. For the snow cover dataset, we create two different partitions; one partition using a canonical geographical grid (which reflects a scenario where the snow cover data provider ignores administrative areas) and one partition that follows the administrative regions (which reflects a scenario where snow cover data are also published by the state governments). The polygons of the grids for the former partition are obtained by dividing the minimum bounding box of Austria into 8 parts in a $4 \times 2$ grid, and the polygons of the states of Austria (used for the partitioning of all three datasets) are obtained from the GADM dataset.

The datasets and the code that we use for partitioning the data is publicly available. The partitioning of a given input dataset according to a set of boundaries is executed as follows: First, we populate each member of the partition with all features that their geometry intersects with the corresponding boundary. Second, we substitute each shape of every member of the partition with its intersection with the corresponding boundary. For all features where their geometry intersects with more than one partitioning boundary, we split the original shape into several parts so that each part fits entirely in a single member of the partition. Finally, we modify the URIs of all resources so that all resources that appear in the same output dataset share a common prefix, which is unique among the prefixes of all datasets of the experiment.

Table 2 and Table 3 illustrate the statistics for the datasets of the experiment; in the table, we group the datasets by type and display statistics about the sum of the group, as-well-as average and standard deviation for each dataset in the group (notice that the datasets are unequal in size due to the large standard deviation for each group). Apart from the statistics, boundary type for each dataset and the number of datasets per group. The number of the snowG datasets is 7 (and not 8, as expected) because the north-west part of the $4 \times 2$ grid does not contain any data due to the shape of Austria.

Each dataset is deployed in a separate GeoSPARQL endpoint. We use the Strabon geospatial RDF store[26] for serving the data. Strabon encapsulates PostGIS for performing spatial operations, and uses a spatial index to optimize query processing time.

**Federations** For the experimental evaluation, we use two possible federation setups for the three available data layers (i.e., administrative, crops, and snow). The first federation setup comprises 27 source endpoints, namely gadm1-9, crops1-9, and snowS1-9; the second one comprises 25 source endpoints, namely gadm1-9, crops1-9, and snowG1-7. In the first setup, we have three datasets for each Austrian state, *i.e.*, all three data layers are split according to the same set of geographical boundaries, while in the second one, the snow cover is divided in a canonical geographical grid, thus we have two data layers that are (by nature) aligned on an uneven geographical split and one that is not aligned.

**Table 2. Information about the datasets used in the evaluation.** For each group of datasets we illustrate the type of thematic data it contains, the boundary of each dataset, and the number of datasets in the group.

| datasets | data type | boundary | #datasets |
|---|---|---|---|
| gadm1-gadm9 | administrative divisions | state polygon | 9 |
| crops1-crops9 | crop types and field boundaries | state polygon | 9 |
| snowS1-snowS9 | snow cover areas | state polygon | 9 |
| snowG1-snowG7 | snow cover areas | 4 × 2 grid | 7 |

**Table 3. Dataset statistics.** For each group of datasets we display statistics about the thematic, geospatial, and both thematic and geospatial (*i.e.*, total) triples in each group. Statistics displayed: sum, average and standard deviation.

| | gadm1-gadm9 | | | crops1-crops9 | | |
|---|---|---|---|---|---|---|
| | total | geospatial | thematic | total | geospatial | thematic |
| sum | 57,087 | 2,231 | 54,856 | 14,056,959 | 2,008,137 | 12,048,822 |
| average | 6,343 | 248 | 6,095 | 1,561,884 | 223,126 | 1,338,758 |
| stdev | 4,888 | 185 | 4,703 | 1,254,534 | 179,219 | 1,075,315 |

| | snowS1-snowS9 | | | snowG1-snowG7 | | |
|---|---|---|---|---|---|---|
| | total | geospatial | thematic | total | geospatial | thematic |
| sum | 331,190 | 66,238 | 264,952 | 335,510 | 67,102 | 268,408 |
| average | 36,799 | 7,360 | 29,439 | 47,930 | 9,586 | 38,344 |
| stddev | 27,349 | 5,470 | 21,879 | 34,272 | 6,854 | 27,418 |

For each federation setup, we set up four Semagrow federators, each with a different source selection configuration. We illustrate all the information about the federations used in the experiment in Table 4. For each federation, we display the source selection method, the number of federated endpoints, details and statistics about the Semagrow metadata used (namely, type of bounding polygon used, metadata size and the number of coordinates that appear in all WKT literals of the `svd:boundingWKT` property), and which datasets from Table 2 used in the federation.

The federators of `thm-27` and `thm-25` use the standard thematic source selection of Semagrow, while the remaining federators use a geospatial source selection on top of it. The difference between the remaining federators is the accuracy of the bounding polygons that the sources are tagged with; in `geo-poly` each source is tagged with the exact polygon that refers to the corresponding areas (*i.e.*, the geographical grid for snowG datasets and the borders of Austrian states for the remaining ones); in `geo-appr` with an approximation of the above polygons a quadtree of height 2; in `geo-mbb` with the minimum bounding box of all shapes that appear in the source. All metadata were created using the Sevod-Scraper tool (see the implementation subsection for more details). The bounding polygons for `geo-mbb-25`, `geo-mbb-27`, `geo-appr-25`, and `geo-appr-27` were calculated by the tool using the minimum bounding box and quadtree-based approximation, while for `geo-poly-25` and `geo-poly-27` we used the exact polygons of the states of Austria and the 4 × 2 grid according to the actual borders of the partitioned datasets. Notice that an increased accuracy leads to an increased metadata size (*i.e.*, even though that the geospatial source selectors use metadata that have the same set of triples, the WKT literals contain a larger set of coordinates).

Regarding the evaluation of a GeoSPARQL query, each Semagrow federation operates as follows: First, federated geospatial joins are evaluated using a bind-join fashion with a filter pushdown optimization. In particular, for each binding of the source query of the left part of the join, Semagrow issues a source query that contains the filter of the geospatial join to the source of the right part of the join. Second, to reduce redundant communication cost, Semagrow group several triple patterns into one source query whenever possible. For instance, triples that appear in a single source are grouped into a single subquery, thus pushing the join operation into the source endpoint.

**Queries** In Table 5, we summarize the queries of the experiment. The query workload is produced by seven query templates (Q1-7); each query template has a single parameter, which is either a WKT literal (Q1-3) or a

**Table 4. Information about the federations used in the evaluation.** For each federation we illustrate the type of the source selector (*i.e.* geospatial or not), number of datasets that appear in the federation, and statistics about the metadata used by the federator.

|  | thm-27 | geo-mbb-27 | geo-appr-27 | geo-poly-27 |
|---|---|---|---|---|
| Source selector | thematic | geospatial | geospatial | geospatial |
| #datasets | 27 | 27 | 27 | 27 |
| Bounding WKT | - | minimum bounding box | approximate shape | exact shape |
| #triples in metadata | 3024 | 3051 | 3051 | 3051 |
| #coordinates in metadata | - | 108 | 1998 | 68736 |
| file size of metadata | 125 KB | 132 KB | 189 KB | 1.8 MB |
|  | **thm-25** | **geo-mbb-25** | **geo-appr-25** | **geo-poly-25** |
| Source selector | thematic | geospatial | geospatial | geospatial |
| #datasets | 25 | 25 | 25 | 25 |
| Bounding WKT | - | minimum bounding box | approximate shape | exact shape |
| #triples in metadata | 2941 | 2959 | 2959 | 2959 |
| #coordinates in metadata | - | 100 | 1360 | 45852 |
| file size of metadata | 123 KB | 127 KB | 165 KB | 1.3 MB |

**Table 5. Queries used in the experiment.**

|     | Parameter | Query |
| --- | --- | --- |
| Q1 | Polygon | Municipalities intersecting a given polygon |
| Q2 | Polygon | Snow-covered potato fields intersecting a given polygon |
| Q3 | Polygon | Potato fields within 5 km from snow cover and intersecting a given polygon |
| Q4 | Municipality name | Snow cover areas within 5 km from a given municipality |
| Q5 | Municipality name | Potato fields within a given municipality |
| Q6 | Municipality name | Snow-covered potato fields within a given municipality |
| Q7 | Municipality name | Potato fields within 5 km from snow cover and within a given municipality |

Municipality name (Q4-7). We generate a set of 100 municipality names and a set of 100 WKT literals; this makes a total of 700 queries.

For the municipality names, we select 100 random municipalities from the GADM shapefile using the PostgreSQL `random()` function. We ignore the municipalities whose names contain characters not in the English alphabet in order to avoid possible string encoding conflicts. For the WKT literals, we create 100 random polygons; We first generate 100 random points within the border of Austria; then, we extend each point by a few meters in each direction, by using the PostGIS `ST_Expand()` function, in order to form rectangles covering approximately an area of 25 square kilometres each. We prune all polygons that are not completely within Austria and repeat the steps above until the random polygons reach 100.

For every query, we define its administrative part as the triple patterns that refer to administrative data (*i.e.,* datasets gadm1-9), its crop part as the triple patterns that refer to crop data (*i.e.,* datasets crops1-9), and its snow part as the triple patterns that refer to snow data (*i.e.,* datasets snowS1-9 or snowG1-7). Q1 comprises only an administrative part, Q2 and Q3 comprise a snow and a crop part, Q4 (respectively Q5) comprises an administrative and a snow (respectively crop) part, Q6 and Q7 contain all three types of parts. Q1-Q3, Q6-7 use the `geof:sfIntersects` function; Q3, Q4, and Q7 use the `geof:distance` function; and finally, Q5-7 use the `geof:sfWithin` function.

For each query template, we also illustrate the number of triple patterns of the query (#tp), the number of geospatial selection filters, *i.e.,* geospatial filters with one free variable (#geoselec), the number of geospatial join filters, *i.e.,* geospatial filters with two free variables (#geojoins), and the relevant data layers for each query template. Notice that the queries that are parameterized with a WKT have geospatial selection filters for each data layer of the query, while the remaning queries do not have geospatial selection filters. Moreover, all queries that make use of two or three data layers (i.e., all queries apart from Q1) have geospatial join filters for combining data from the corresponding layers.

Finally, in Table 6 we illustrate statistics about the queries in the evaluation. In particular, we illustrate the average number of results (#r) of all queries that belong to each query template. Notice that the queries return a small number of results (*e.g.,* in Q6, for several municipalities the query is expected to return no results). This fact does not necessarily mean that the setup is not challenging enough. Using large-scale queries (and datasets) would be needed to evaluate the efficiency of the federated query execution engine of the federator. But since the subject of this evaluation is the source selection engine, discussing query execution efficiency would be a digression. If anything, using larger datasets and queries would make the source selection time overheads look even smaller, relatively, so our evaluation setup is actually stricter on ourselves than an evaluation on larger datasets.

**Experiment deployment and execution** We use a Kubernetes 1.14 cluster with 1 master node and 8 worker nodes with a total if 120 cores and 264GB RAM. Experiment deployment and execution is done through the KOBE benchmarking engine[27], and the KOBE configurations for reproducing the experiments are publicly available.

**Table 6. Query statistics.**

|    | # triple patterns | # geospatial selections | # geospatial joins | thematic data layers used | # results |
|----|-------------------|-------------------------|--------------------|---------------------------|-----------|
| Q1 | 6                 | 1                       | 0                  | gadm                      | 3.7       |
| Q2 | 10                | 2                       | 1                  | crops, snow               | 2.1       |
| Q3 | 10                | 2                       | 1                  | crops, snow               | 15.6      |
| Q4 | 9                 | 0                       | 1                  | gadm, snow                | 12.5      |
| Q5 | 9                 | 0                       | 1                  | gadm, crops               | 9.7       |
| Q6 | 14                | 0                       | 3                  | gadm, crops, snow         | 0.5       |
| Q7 | 14                | 0                       | 3                  | gadm, crops, snow         | 6.7       |

## Experimental results

In the following, we present the experimental results. We first focus on each phase of federated query processing separately, and then we discuss total query processing as a whole.

**Evaluation metrics** All queries are decomposed successfully and for every query a correct execution plan is produced. However, in some queries (*e.g.,* in some query instances of Q4 and Q7) the query execution phase evokes an error, and in these situations the federator returns no answer.

The experimental results are summarized in Table 7, Table 8, Table 9, Table 10, Table 11, and Table 12. For each query template of Table 5 and for each federation of Table 4, we display the following evaluation metrics: to evaluate the efficiency of the query execution plan we display the average query execution time of all successful queries (Table 11) and the error rate (Table 10), *i.e.,* the number of the unsuccessful queries over the number of all queries in the template; to check the efficiency of the other parts of query processing we display the average source selection time (Table 7) and planning time (Table 9); to check if the source selection time overheads are recovered by reduced planning and execution time, we display the time differences between each geospatial federation with its corresponding thematic one (Table 12); to evaluate the efficiency of the pruning of each source selection, we display the average number of sources that are accessed during the evaluation (Table 8); and finally, to check if any source selector achieves optimal pruning, we include the average size of the optimal source set (`opt-27` and `opt-25` columns of Table 8).

Regarding the time measurements shown in Table 7, Table 9, Table 11, and Table 12, apart from the average value, we include its standard derivation (displayed in parentheses). Moreover, regarding the average number of sources in Table 8, we include in parentheses the minimum and maximum number of sources.

**Comparison of source selection times** In the following, we focus on the time overheads of the geospatial source selector. Thus, we will compare the federations of the experiment according to source selection time (Table 7).

We observe that the source selectors of `thm-27` and `thm-25` (in short `thm`) are the fastest ones; then we have `geo-mbb-27` and `geo-mbb-25` (in short `geo-mbb`); then we have `geo-appr-27` and `geo-appr-25` (in short `geo-appr`); and finally we have `geo-poly-27` and `geo-poly-25` (in short `geo-poly`). This happens due to two main reasons. First, the geospatial source selector (*i.e.,* that of `geo-mbb`, `geo-appr` and `geo-poly`) wraps the thematic source selector of Semagrow (*i.e.,* that of `thm`), which explains why `thm` is the fastest of all. Second, the sources in `geo-poly` are annotated with polygons, which are more complex shapes than the approximated shapes in `geo-appr`, which are, in turn, more complex shapes than the bounding boxes in `geo-mbb`. Thus, the boundary comparisons performed by the geospatial source selection are slower in `geo-poly`. This difference is more pronounced in Q3 and Q7, which include three geospatial filters and a within-distance operation (using the `geof:distance` function), which is computationally costlier than containment and intersection operations.

We observe that, in general, the source selection process is faster in the federations with 25 endpoints (*e.g.,* compare the source selection time for the queries for `geo-poly-25` with `geo-poly-27`). This happens not only because in the 27-dataset setup we have two additional endpoints, but mainly because the snow data in the 25-dataset setup is partitioned using a canonical grid. Therefore, the boundary annotations of the snow datasets are rectangles not only in `geo-mbb-25` (as expected), but in `geo-poly-25` and `geo-appr-25` as well. As a result, the geospatial computations performed by the source selector for identifying irrelevant snow sources is much faster in the 25-dataset setup.

**Table 7. Source Selection time (sec): Average (and standard deviation) of 100 instances per query template (Q1–Q7).** We display metrics for each federation (*i.e.*, `geo-poly-27`, `geo-poly-25`, `geo-appr-27`,etc.) of the evaluation.

|    | geo-poly-27 | | geo-appr-27 | | geo-mbb-27 | | thm-27 | |
|----|------|--------|------|--------|------|--------|------|--------|
| Q1 | 0.10 | (0.02) | 0.14 | (0.02) | 0.13 | (0.02) | 0.08 | (0.01) |
| Q2 | 0.87 | (0.06) | 0.26 | (0.02) | 0.22 | (0.02) | 0.13 | (0.02) |
| Q3 | 8.28 | (0.23) | 0.74 | (0.09) | 0.22 | (0.03) | 0.14 | (0.02) |
| Q4 | 1.76 | (0.14) | 0.37 | (0.16) | 0.21 | (0.08) | 0.17 | (0.07) |
| Q5 | 0.42 | (0.08) | 0.21 | (0.02) | 0.20 | (0.03) | 0.26 | (0.13) |
| Q6 | 1.57 | (0.10) | 0.37 | (0.09) | 0.30 | (0.11) | 0.22 | (0.07) |
| Q7 | 8.53 | (0.24) | 1.11 | (0.34) | 0.42 | (0.20) | 0.39 | (0.26) |
|    | geo-poly-25 | | geo-appr-25 | | geo-mbb-25 | | thm-25 | |
| Q1 | 0.16 | (0.02) | 0.14 | (0.02) | 0.13 | (0.02) | 0.08 | (0.01) |
| Q2 | 0.46 | (0.03) | 0.14 | (0.02) | 0.20 | (0.02) | 0.13 | (0.02) |
| Q3 | 0.43 | (0.04) | 0.21 | (0.03) | 0.19 | (0.02) | 0.14 | (0.08) |
| Q4 | 0.36 | (0.09) | 0.19 | (0.04) | 0.19 | (0.07) | 0.16 | (0.06) |
| Q5 | 0.33 | (0.08) | 0.29 | (0.07) | 0.19 | (0.02) | 0.22 | (0.08) |
| Q6 | 0.76 | (0.07) | 0.29 | (0.11) | 0.28 | (0.09) | 0.18 | (0.04) |
| Q7 | 0.83 | (0.14) | 0.39 | (0.14) | 0.35 | (0.14) | 0.41 | (0.34) |

**Table 8. Source Selection pruning: number of sources selected by the different source selection methods, average (min- imum and maximmum) over 100 query instances per query template (Q1–Q7).** We display metrics for each federation (*i.e.*, `geo-poly-27`, `geo-poly-25`, `geo-appr-27`,etc.) of the evaluation, as well as the optimal ones (*i.e.*, `opt-27`, `opt-25`).

|    | opt-27 | | geo-poly-27 | | geo-appr-27 | | geo-mbb-27 | | thm-27 |
|----|------|-------|------|-------|------|--------|-------|---------|------|
| Q1 | 1.12 | (1–2) | 1.12 | (1–2) | 1.32 | (1–3) | 1.62 | (1–3) | 9 |
| Q2 | 2.24 | (2–4) | 2.24 | (2–4) | 2.64 | (2–6) | 3.24 | (2–6) | 18 |
| Q3 | 2.24 | (2–4) | 2.24 | (2–4) | 2.64 | (2–6) | 3.24 | (2–6) | 18 |
| Q4 | 2.25 | (2–4) | 5.48 | (3–7) | 5.48 | (3–7) | 5.90 | (3–7) | 10 |
| Q5 | 2.00 | (2–2) | 2.00 | (2–2) | 5.48 | (3–7) | 5.82 | (3–7) | 10 |
| Q6 | 3.00 | (3–3) | 3.00 | (3–3) | 9.96 | (5–13) | 10.64 | (5–13) | 19 |
| Q7 | 3.24 | (3–5) | 6.48 | (4–8) | 9.96 | (5–13) | 10.64 | (5–13) | 19 |
|    | opt-25 | | geo-poly-25 | | geo-appr-25 | | geo-mbb-25 | | thm-25 |
| Q1 | 1.12 | (1–2) | 1.12 | (1–2) | 1.32 | (1–3) | 1.62 | (1–3) | 9 |
| Q2 | 2.18 | (2–4) | 2.18 | (2–4) | 2.38 | (2–4) | 2.68 | (2–5) | 16 |
| Q3 | 2.18 | (2–4) | 2.18 | (2–4) | 2.38 | (2–4) | 2.68 | (2–5) | 16 |
| Q4 | 2.14 | (2–3) | 4.34 | (2–5) | 4.59 | (2–5) | 4.59 | (2–5) | 8 |
| Q5 | 2.00 | (2–2) | 2.00 | (2–2) | 5.48 | (3–7) | 5.82 | (3–7) | 10 |
| Q6 | 3.18 | (2–4) | 4.97 | (3–6) | 8.82 | (4–11) | 9.41 | (4–11) | 17 |
| Q7 | 3.18 | (3–5) | 4.97 | (3–6) | 8.82 | (4–11) | 9.41 | (4–11) | 17 |

**Table 9. Query planning time (sec): Average (and standard deviation) of 100 instances per query template (Q1–Q7).** We display metrics for each federation (*i.e.*, `geo-poly-27`, `geo-poly-27`, `geo-appr-27`,etc.) of the evaluation.

| | `geo-poly-27` | | `geo-appr-27` | | `geo-mbb-27` | | `thm-27` | |
|---|---|---|---|---|---|---|---|---|
| Q1 | 0.02 | (0.01) | 0.03 | (0.01) | 0.03 | (0.01) | 0.04 | (0.01) |
| Q2 | 0.28 | (0.04) | 0.30 | (0.03) | 0.31 | (0.03) | 0.39 | (0.04) |
| Q3 | 0.26 | (0.04) | 0.24 | (0.04) | 0.26 | (0.03) | 0.38 | (0.04) |
| Q4 | 0.10 | (0.02) | 0.15 | (0.09) | 0.13 | (0.03) | 0.19 | (0.07) |
| Q5 | 0.12 | (0.03) | 0.11 | (0.02) | 0.12 | (0.01) | 0.15 | (0.02) |
| Q6 | 13.56 | (0.29) | 14.48 | (0.44) | 14.33 | (0.40) | 16.00 | (0.35) |
| Q7 | 13.69 | (0.31) | 14.55 | (1.37) | 14.81 | (2.61) | 16.55 | (3.95) |
| | `geo-poly-25` | | `geo-appr-25` | | `geo-mbb-25` | | `thm-25` | |
| Q1 | 0.03 | (0.01) | 0.03 | (0.01) | 0.03 | (0.01) | 0.05 | (0.01) |
| Q2 | 0.27 | (0.05) | 0.21 | (0.02) | 0.31 | (0.02) | 0.39 | (0.04) |
| Q3 | 0.25 | (0.05) | 0.26 | (0.03) | 0.26 | (0.03) | 0.38 | (0.05) |
| Q4 | 0.13 | (0.07) | 0.14 | (0.10) | 0.14 | (0.07) | 0.19 | (0.13) |
| Q5 | 0.13 | (0.02) | 0.12 | (0.02) | 0.12 | (0.01) | 0.15 | (0.02) |
| Q6 | 14.30 | (0.49) | 14.70 | (0.48) | 14.27 | (0.46) | 15.56 | (0.29) |
| Q7 | 13.66 | (0.28) | 14.10 | (0.46) | 14.64 | (2.43) | 17.24 | (6.36) |

**Table 10. Error rate: Number of errors occured, divided by the total number of queries of each query template (Q1–Q7).** We display metrics for each federation (*i.e.*, `geo-poly-27`, `geo-poly-25`, `geo-appr-27`,etc.) of the evaluation.

| | `geo-poly-27` | `geo-appr-27` | `geo-mbb-27` | `thm-27` | `geo-poly-25` | `geo-appr-25` | `geo-mbb-25` | `thm-25` |
|---|---|---|---|---|---|---|---|---|
| Q1 | - | - | - | - | - | - | - | - |
| Q2 | - | - | - | - | - | - | - | - |
| Q3 | - | - | - | 0.1 | - | - | - | - |
| Q4 | 0.1 | 0.1 | 0.1 | 0.1 | 0.1 | 0.1 | 0.1 | 0.1 |
| Q5 | - | - | - | - | - | - | - | - |
| Q6 | - | 0.7 | 0.7 | 0.8 | - | 0.7 | 0.7 | 0.1 |
| Q7 | 0.1 | 0.9 | 0.9 | 0.9 | 0.1 | 0.9 | 0.9 | 0.9 |

**Table 11. Query execution time (sec): Average (and standard deviation) of 100 instances per query template (Q1–Q7).** We display metrics for each federation (*i.e.*, `geo-poly-27`, `geo-poly-25`, `geo-appr-27`, etc.) of the evaluation.

| | geo-poly-27 | | geo-appr-27 | | geo-mbb-27 | | thm-27 | |
|---|---|---|---|---|---|---|---|---|
| Q1 | 0.06 | (0.05) | 0.04 | (0.03) | 0.04 | (0.04) | 0.06 | (0.06) |
| Q2 | 0.05 | (0.05) | 0.04 | (0.02) | 0.04 | (0.02) | 0.17 | (1.30) |
| Q3 | 0.21 | (1.20) | 0.17 | (1.12) | 0.17 | (1.15) | 0.06 | (0.25) |
| Q4 | 6.87 | (4.18) | 6.47 | (3.49) | 6.81 | (3.55) | 8.52 | (4.96) |
| Q5 | 0.13 | (0.08) | 0.08 | (0.03) | 0.09 | (0.06) | 0.12 | (0.09) |
| Q6 | 0.11 | (0.05) | 2.20 | (2.21) | 2.18 | (2.39) | 2.30 | (1.47) |
| Q7 | 23.19 | (67.86) | 4.14 | (1.63) | 4.43 | (1.06) | 6.64 | (2.30) |
| | geo-poly-25 | | geo-appr-25 | | geo-mbb-25 | | thm-25 | |
| Q1 | 0.05 | (0.05) | 0.04 | (0.01) | 0.04 | (0.01) | 0.05 | (0.02) |
| Q2 | 0.05 | (0.05) | 0.04 | (0.03) | 0.04 | (0.02) | 0.30 | (1.32) |
| Q3 | 0.19 | (1.15) | 0.17 | (1.09) | 0.17 | (1.13) | 2.52 | (10.28) |
| Q4 | 7.35 | (4.73) | 7.33 | (4.70) | 7.43 | (4.77) | 8.81 | (5.27) |
| Q5 | 0.26 | (0.41) | 0.11 | (0.05) | 0.08 | (0.03) | 0.10 | (0.04) |
| Q6 | 0.77 | (0.83) | 1.37 | (1.10) | 1.52 | (1.50) | 1.46 | (1.04) |
| Q7 | 19.44 | (57.78) | 3.10 | (2.33) | 3.07 | (2.37) | 38.69 | (70.86) |

**Table 12. Time overhead of the geospatial source selection: Average (and standard deviation) of the difference in total query processing time (in seconds) of each geospatial federation minus the time of its corresponding thematic one, over the successful query instances of query template Q1 to Q5.** A negative measurement indicates that the geospatial source selection overheads are recovered by faster query planning and execution. Q6 and Q7 are missing from the table since most queries in `geo-appr,` `geo-mbb`, and thm evoke errors during query execution phase.

| | geo-poly-27 | | geo-poly-27 | | geo-mbb-27 | |
|---|---|---|---|---|---|---|
| Q1 | −0.01 | (0.0000) | +0.03 | (0.0000) | +0.03 | (0.0001) |
| Q2 | +0.51 | (0.0013) | −0.10 | (0.0013) | −0.12 | (0.0013) |
| Q3 | +8.03 | (0.0004) | +0.45 | (0.0003) | −0.05 | (0.0002) |
| Q4 | +0.23 | (0.0017) | −1.28 | (0.0018) | −1.20 | (0.0017) |
| Q5 | +0.14 | (0.0002) | −0.13 | (0.0002) | −0.12 | (0.0002) |
| | geo-poly-25 | | geo-appr-25 | | geo-mbb-25 | |
| Q1 | +0.05 | (0.0000) | +0.03 | (0.0000) | +0.01 | (0.0000) |
| Q2 | −0.05 | (0.0013) | −0.43 | (0.0013) | −0.27 | (0.0013) |
| Q3 | −1.16 | (0.0027) | −1.39 | (0.0027) | −1.39 | (0.0027) |
| Q4 | −0.71 | (0.0028) | −1.01 | (0.0028) | −0.85 | (0.0029) |
| Q5 | +0.03 | (0.0001) | +0.04 | (0.0001) | −0.09 | (0.0001) |

To sum up, we notice that source selection time depends on the complexity of the bounding polygon annotations of the sources; in other words, higher accuracy leads to slower source selection.

**Comparison of source selection pruning** In the following, we focus on the precision of the pruning of each source selector (Table 8). In particular, we compare the number of sources of each source selector with those of the other source selectors and with the optimal ones.

We observe that the thematic source selector keeps many irrelevant sources in the query plan. The source selector of thm exploits the thematic information (*i.e.,* properties and URI-prefixes) of the sources and assigns the administrative (respectively crop, snow) part of the query to the administrative (resp. crop, snow) sources. Moreover, in Q4-7, the administrative part of the query is further restricted to a single administrative source, because the pattern that specifies the name of the municipality appears in only a single administrative source. This explains why, for example, thm-27 keeps 19 (*i.e.,* 9 crop sources, 9 snow sources, and 1 administrative source) and not all 27 sources for all queries in the template. However, as expected, we will show that the geospatial selectors of the remaining federations achieve better pruning by exploiting the geospatial knowledge of the sources.

Regarding the geospatial selectors, we make three observations: First, we notice that the accuracy of the source selector increases as the accuracy of the source metadata increases. In particular, geo-poly is more precise than geo-appr, and geo-appr is more precise than geo-mbb. Second, we notice that the optimal pruning can be achieved only by geo-poly in Q1-3, Q5 (and also in Q4 only for the 27-dataset setup). Finally, the average number of sources for geo-appr and geo-mbb tend to be lower in Q1-3 than in Q4-7. In the remaining paragraphs we will try to discuss these observations. Q1-3 have geospatial selection filters, parameterized with a fixed polygon; thus, the geospatial selector operates by pruning all sources that are irrelevant according to the given query polygon. Since the less accurate geospatial summaries in geo-appr and geo-mbb are larger than the actual dataset boundaries of the sources, we can have a situation where the query polygon is disjoint from a data source but not disjoint from its bounding polygon annotation. This explains the optimal pruning for geo-poly (where the annotations are the exact boundaries of the sources). In the remaining geospatial federations, the source selection returns more sources, because there are cases where the parameterized polygon is contained in the approximated shape (for geo-appr) or in the minimum bounding box (for geo-mbb) of a neighbor source. This explains why geo-appr is equally or more specific than geo-mbb.

Q4-7 contain only geospatial join filters; therefore, the geospatial selector operates as follows; first, similarly to thm, it restricts the administrative part of the query in the source of the state where the municipality belongs to; then, it tries to prune all irrelevant crop and snow sources according to the boundary annotation of this administrative source and the geospatial filters of the query. As previously, we observe that accurate source descriptions can lead to more precise source selection. For instance, regarding Q5, geo-mbb (resp. geo-appr) prunes all crop sources that their bounding box (resp. approximated shape) is disjoint from the bounding box (resp. approximated shape) of the state of interest, while geo-poly, being more accurate, does better by keeping only the crop sources that refer this state, because the source boundaries do not overlap. This explains the optimal pruning of geo-poly for Q5.

Q4-7 present two additional challenges in source selection, which are either non-present or non-important in Q1-3. First, Q4 and Q8 contain a within-distance federated join operation, but unlike Q3, the shapes of interest do not intersect with a given polygon in the query. In such operations, the geospatial selector cannot achieve optimal pruning even in geo-poly. To give an example, consider Q4 and assume that the given municipality appears towards the center of the state. Since the exact geometric shape of the municipality will be discovered only during query execution, the source selector cannot exclude the possibility of its position being towards the border, thus keeping all the neighboring snow sources that may contain relevant data within 5km from the border of the state. Second, an overestimation of the set of sources can appear when the geographical partitions between the data to be geospatially joined are unaligned. Consider Q6; since in the 27-dataset setup all data layer partitions are geographically aligned (each source refers to a specific Austrian state), geo-poly-27 achieves optimal pruning (*i.e.,* the source selector keeps the sources that refer to the state where the municipality belongs to). In contrast, since in the 25-dataset setup the snow data partition is not aligned with the other layers, geo-poly-25 keeps some irrelevant neighboring snow sources (*i.e.,* those who intersect the state that belongs to the municipality but not the municipality itself) and thus does not achieve optimal pruning.

To sum up, we notice that the precision of the pruning by the geospatial source selector depends on the accuracy of the bounding polygon annotations of the sources. We observe that using the exact polygons of the sources could lead us to optimal pruning. Finally, we notice that in queries with WKT parameters (Q1-3) the geospatial source selectors tend to achieve a better pruning, even when using approximated shapes instead of exact polygons.

**Comparison of query planning and execution** In the following, we discuss the effect of geospatial source selection on query planning and query execution phases of federated query processing. In particular, we compare the query planning times (Table 9), the query execution times (Table 11), and the error rates (Table 10) of each federation of the experiment.

Regarding query planning time, we observe that, in general, `geo-poly` is the fastest; then it comes `geo-appr`; then we have `geo-mbb`; and finally `thm` is the slowest. This behaviour happens because having a large number of sources requires the construction of a larger query plan, which clearly affects the time for producing it; this is highlighted in Q6 and Q7 which have 14 triple patterns (the number of triple patterns of each query is shown in Table 6).

Regarding query execution, notice that only for some query templates we obtain a complete evaluation of all queries in the template. For instance, ~90% of the queries of Q7 fail to be processed by `thm-27` due to errors in the execution phase (*i.e.,* the error rate in Table 10 is equal to 0.9). These errors occur when a federator issues a huge workload of source queries to the endpoints, and as a result, the sources are not able to serve all these requests. Therefore, in order to compare two query executions, we should consider both their query execution times and their error rates. For instance, consider again Q7; the query execution of `thm-27` is faster than that of `geo-poly-27`, but the error rate of `geo-poly-27` is much lower than that of `thm-27`. Thus, we argue that `geo-poly-27` is more effective than `thm-27` for Q7, because we believe that having more but slower successful query runs is a more important characteristic (recall that the average execution time refers only to successful query runs).

The number of source queries in the execution plan affects not only the completion of the execution but the execution time as well; having more sources in the plan means that more source queries are issued by the federator to the source endpoints. Consider, for instance, Q2 and Q3; in both cases, the query execution of `geo-poly-25` is faster than that of `thm-25` by one order of magnitude; `geo-poly-25` consults one (or in some cases two) snow datasets, while `thm-25` consults all 7 snow datasets. Even though Semagrow manages to execute many queries in parallel, the duration of the query execution has to be as slow as the slowest source. By having a smaller the set of sources, the query executor avoids issuing queries to irrelevant larger datasets if they contain irrelevant results. Moreover, the time difference in query execution is more pronounced in queries that contain within-distance operations (*e.g.,* in Q3), because the source endpoints use spatial indexes, hence geospatial queries that contain standard spatial relations (*e.g.,* Q2) are executed faster.

The above discussion suggests that, according to the effectiveness of their query execution (which is based both on error rate and query execution time), the federators are to be ordered as follows: `geo-poly`, `geo-appr`, `geo-mbb`, and finally `thm`; the only exception being Q6 in the 25-dataset setup. In this sole case, `geo-poly-25` is better than `thm-25`, but `thm-25` has lower error rate than `geo-appr-25` and `geo-mbb-25`. This final observation indicates that, even though our geospatial source selector can provide a faster query processing, it seems that in order to achieve better performance in geospatial scenarios, the remaining components of federated query processing should be extended with geospatial-specific optimizations as well.

To sum up, we observe that higher accuracy in geospatial source annotations (which results to a lower number of sources per query) could help by reducing the query planning time and the number of source queries issued by the federator, thus increasing the effectiveness of query execution.

**Comparison of overall query processing time** The discussion so far indicates that using a geospatial selector provides a positive impact on query planning and execution time. However, as the accuracy of the bounding polygons of the federated sources increases, source selection becomes slower, especially when using the exact polygons (as in `geo-poly`). The question that arises is whether the time overhead of the use of exact boundaries in source selection can be recovered by the remaining phases of query processing.

In Table 12, we draw a comparison between the time overheads of the geospatial source selectors; among the query instances that succeed in all eight federations, we show the difference of the total query processing time of `geo-poly-27` (resp. `geo-appr-27`, `geo-mbb-27`) minus the total query processing time of `thm-27`; then, the same for the 25-dataset setup; and finally, we report the average (and standard deviation) for each query template. We leave out Q6 and Q7 because in both query templates less than 5 instances succeed in all 6 federations; in this case, we will compare the federations with respect to the error rate (Table 10).

Q1 is the easiest query of the experiment (it contains a single data layer and one geospatial selection filter, *i.e.,* a filter that contains a spatial relation in which one of the two parameters is a WKT value). Thus, all federators perform equally in Q1 (*i.e.,* all time differences are less than 0.05 seconds). In contrast, Q6 and Q7 are the most difficult queries of the experiment (they contain 3 data layers, three geospatial joins, and no WKT literals appear in the query body). Since most query instances of Q6 and Q7 fail to be processed, we compare the federations w.r.t. Table 10; we note that `geo-poly` performs the best since it minimizes the error rate.

The remaining queries (*i.e.,* Q2-5) are somewhere in between Q1 and Q6-7 in terms of difficulty; this fact makes them easier to be processed by all federators of the experiment. In Table 12 we notice that `geo-mbb` and `geo-appr` outperform `thm` and `geo-poly` (overheads are smaller or similar). Regarding the comparison between `geo-mbb` and `geo-appr` though, we observe that `geo-mbb` is better in the 27-dataset setup, while `geo-appr` is better in the 25-dataset setup. This happens because the source selection cost in the 27-dataset setup is much higher than that of the 25-dataset setup (Table 7). Thus, in the former setup, only `geo-mbb-27` can benefit from the reduction in query planning and execution times, while in the second one, the drop in planning and execution time of `geo-appr-25` is greater than its source selection overhead.

To sum up, we observe that for difficult queries (such as queries that contain more than one geospatial join and no WKT literals in the query body), precise bounding polygons should be preferred, because otherwise we may face a computationally intensive query execution. In contrast, the use of less accurate descriptions will suffice if we consider simpler queries. However, it appears that no size fits all; for the setup that the partitions are unaligned, we benefit from the higher accuracy of the approximated shapes since one layer is already a geographical grid; while for the other setup the minimum bounding boxes are effective since all data layers are fully aligned according to the same administrative regions.

## Related work

Federated querying is specified in the SPARQL 1.1 standard[28], and also its GeoSPARQL extension, but the standard does not require that sources are prudently selected. The standard allows the query processor to flood all sources of the federation with all triple patterns unless the query author uses the SERVICE keyword to explicitly restrict the source or sources for some of the patterns. Several GeoSPARQL implementations exist that support federated querying without source selection[29-32].

Despite the rich literature on thematic source selection discussed in the background section, work on federated geospatial query processing is very sparse, in the Semantic Web community, the geographical information systems community, and the wider databases community. Recent studies[5] find that there is no mature federated GeoSPARQL query processing system. Recent work on data integration methods cites systems that collect and integrate distributed geospatial data into a single store as well dynamic federation of non-geospatial data sources, but also does not include systems that are both federated and support geospatial operations[33]. Other previous works on federating GeoSPARQL endpoints apply state-of-the-art thematic source selection, but without any method for eliminate candidate sources based on their geospatial extent[34,35]. Other works on federated geospatial processing do not deal with the concept of source selection at all, because they target joining data of different geospatial formats, such as joins between vector and raster data[36,37].

A related system is SkyQuery[38], a federation of astronomy databases. SkyQuery optimizes the execution of SQL-like queries based on metadata similar in nature to the VoID dataset statistics, provided by the individual databases when registering to the federation. Cross match queries, in particular, use containment constraints to retrieve objects corresponding to the same astronomical body allowing for some error in the measurements. The constraint is satisfied by looking up an index of spherical triangles, operating analogously to our bounding WKTs: for each triangle the index points to all databases that hold objects within the triangle. In fact, the triangles that make up the SkyQuery index are organized as a containment hierarchy, analogously to how R-trees are used in geospatial database indexes[38]. This higher level of detail (by comparison to our system) allows SkyQuery to use its index not only for source selection but to fully optimize query execution.

Zimmermann *et al.*[39] propose to use R-trees and Quadtrees as index structures across multiple spatial databases to reduce the query forwarding traffic. Each archive maintains a copy of such a global index with the minimum bounding rectangle (MBR) of the dataset of each archive. The archives can determine through their local copy of the global index which of the other archives might have relevant data (*i.e.* whose MBR overlaps/intersects with the query rectangle). The query routing does not contain all the servers in a distributed spatial database environment in order to obtain the query results. The query is only forwarded to archives with potentially relevant data, decreasing the inter-node message traffic significantly. The infrastructure used by Zimmermann *et al.*[39] to orchestrate query routing cannot be directly mapped to GeoSPARQL endpoints as they currently stand, so a major effort would be needed to extend Semantic Web infrastructure. Our method, by contrast, only requires a light-weight extension to the VoID vocabulary already used to summarize datasets.

Similarly, Tang *et al.*[40] introduce a framework for integrated queries for geospatial data services. They maintain an R-tree index with MBRs of the services' spatial extent and prune services that cannot possibly contribute to k-nearest neighbours queries based on the maximum and minimum distance between the query shape and these MBRs. Although the core idea is similar, Tang *et al.*[40] only support k-NN operations and their system does not present to the user a complete federated querying endpoint. By contrast, we demonstrate complete support for GeoSPARQL querying.

## Conclusions and future work

We presented a source selection method that combines the thematic source selection typically used in federated query processing with an additional data source filtering based on the bounding polygon (expressed as a WKT value literal) that summarizes the geospatial extent of all resources in a data source. The prototype implementation of our method is provided as open source, integrated with the Semagrow federated GeoSPARQL processor. The data and queries used in the experiments are also published as the `geofedbench` benchmark of the KOBE benchmarking environment[41].

We explored three alternative bounding WKTs of varying accuracy. More complex bounding WKTs lead to slower source selection run-times, but also to more precise exclusion of unneeded sources, so that the sources in the federation are not burdened by pointless querying, which may inundate the sources with queries to the point of failure. Experimental results show that our method has substantial positive impact in overall query processing time as well. In particular, the source selection run-time is (partially or fully) recovered by shorter planning and execution run-time; and more precise selection makes the federation engine more prudent with the web resources it consumes.

Regarding the accuracy of the bounding WKTs used in our evaluation, we experimented with the (exact) minimum bounding polygons, the minimum bounding boxes, and approximated polygons that fall in between the above two extremes in terms of accuracy. The former two summaries are straightforward to compute; but there are many other ways of balancing the trade-off between summary size (which relates to source selection time) and summary accuracy (which relates to source selection precision). As future work, we are planning to continue our research on finding the proper geospatial summarization techniques for maximizing the performance of the source selection. One possible direction would be to use a convex-hull algorithm to compute the optimal bounding WKT with a fixed user-defined number of edges. This will allow us to make the exact place where the system sits on the bounding box vs. bounding polygon trade-off user-configurable.

What we can safely confirm from our experiments is that there is value in geospatial source selection because the computational overheads are compensated by the greatly reduced query execution time, benefiting both the server and the client. However, regardless of whether the inventory of possible summaries is expanded or not, already for the three summary types presented here there is no clear-cut conclusion about which one works best: Simpler summaries should be preferred for computationally easy queries and finer summaries should be preferred for more demanding queries with long query processing times. Given this, we are planning to explore the deeper integration of geospatial summaries into all federated query processing phases, most critically query planning. Since query planning already estimates query execution time in order to drive optimizations, it start with a rough geospatial summary as a default and estimate whether finer summaries should be retrieved and used for re-planning and source selection.

## Ethics and consent

Ethical approval and consent were not required.

## Data availability

### Source data

As discussed in our experimental setup, our evaluation depends on administrative, crop type, and snow cover publicly available data (Table 2 and Table 3 and Experimental setup subsection).

The KOBE experimental setup used to carry out these experiments is configured to fetch all datasets from the following repository: http://rdf.iit.demokritos.gr/dumps/gss (DOI: https://doi.org/10.5281/zenodo.6340417)

These have been prepared by pre-processing the following:

- Austria administrative areas, https://gadm.org/maps/AUT.html

- INVEKOS, https://www.data.gv.at/katalog/dataset/f7691988-e57c-4ee9-bbd0-e361d3811641

- Snow cover data from the Extreme Earth project's Food Security use case, http://earthanalytics.eu/food-security-use-case.html

## Software availability

The software that refers to this work is publicly available.

Implementation of our source selection: https://github.com/semagrow/semagrow/releases/tag/2.2.0-gssbench

Code to conduct the experiments: https://github.com/semagrow/benchmark-geofedbench/releases/tag/1.0.0

Archived code at time of publication: DOI: 10.5281/zenodo.6341487

License: Apache License, Version 2.0

# Open Peer Review

## Current Peer Review Status:

---

**Version 2**

Reviewer Report 21 October 2022

https://doi.org/10.21956/openreseurope.16399.r30243

**Markus Stocker**

TIB - Leibniz Information Centre for Science and Technology, Hannover, Germany

The authors have addressed reviewer comments reasonably well. I think we can debate on the impact of the work and the practical relevance of the technologies for much longer, so perhaps this discussion is material for another paper.

***Competing Interests:*** No competing interests were disclosed.

***Reviewer Expertise:*** My areas of expertise include semantic web technologies as well as geospatial data and query processing.

**I confirm that I have read this submission and believe that I have an appropriate level of expertise to confirm that it is of an acceptable scientific standard.**

Reviewer Report 20 October 2022

https://doi.org/10.21956/openreseurope.16399.r30242

**Ozgu Can**

Department of Computer Engineering, Faculty of Engineering, Ege University, Bornova, Turkey

My comments have been mostly resolved. I suggest authors to add the following issues regarding the current version of the manuscript before the final publication:
- Figure 4 is very superficial to demonstrate the architecture. The figure could be improved.

○ Authors claim that real-life use cases are limited. Thus, the limitations of the study should be discussed in the conclusion.

**Competing Interests:** No competing interests were disclosed.

**Reviewer Expertise:** Semantic Web, Knowledge Management, Ontology Engineering, Security, Privacy,

**I confirm that I have read this submission and believe that I have an appropriate level of expertise to confirm that it is of an acceptable scientific standard.**

---

**Version 1**

Reviewer Report 20 July 2022

**Markus Stocker**

TIB - Leibniz Information Centre for Science and Technology, Hannover, Germany

The authors propose an approach that enables GeoSPARQL queries to determine which sources to geospatial data to query, in a federated setting. The work is technically sound and evaluated sufficiently well. However, I personally found it very hard to read. Aside from occasional typos, sentences are ofter far too long and complicated. I think authors should substantially revise the manuscript and improve readability.

The authors claim that federated query processing is urgently needed to achieve "full" integration of geospatial data in the Semantic Web. Not sure I agree, since spatial querying in the Semantic Web is not obvious even on standalone databases. More generally, how much of a problem is it actually? The Semantic Web has been loosing ground for some time; the adoption of RDF and GeoSPARQL in (research) infrastructure is rare and the prospects not exactly great (in my opinion). On the other hand, there is solid GIS tech out there (PostGIS, JTS Topology Suite, etc.) that do an excellent job and seem to do just fine without RDF, GeoSPARQL, etc. Hence, while I appreciate the academic exercise of developing source selectors for GeoSPARQL I do wonder how much of a problem it actually is.

For an academic work, I am afraid it does not cover well enough the literature. The authors limit the review to the Semantic Web area, which is not adequate because more related work has been done outside the Semantic Web area. Moreover, foundational work such as RCC are not reviewed adequately. Even within the Semantic Web area, there is related work that has not been considered. My suggestion is that the manuscript should have 2-3 times as many citations.

In Definition 2, the authors suggest that the domain of the newly introduced property is void:Dataset. What is the implication of this restriction? Does the approach only work for individuals typed void:Dataset. What about all the other applicable types, say, qb:DataSet or dcat:Dataset? The choice seems to be very restrictive.

Most of the results are obvious, expected and thus unsurprising. For instance, it does not surprise that the thematic source selector is faster than geospatial source selectors, since it is computationally much cheaper to match some URIs than to perform polygon computations. The same can be said about source selection pruning: Of course the selector does a better job the more precise the information available.

The overall conclusion seems that "difficult queries" rely on precise bounding polygons while less accurate descriptions are sufficient for "simpler queries". As the authors phrase it: "no size fits all." What does this mean in practice? Do all datasets have to be described with both precise and less-precise bounding boxes, so that the system can choose based on query complexity? And if so how would this concretely work? What is the recommendation?

As a minor remark, I suggest to move the result tables closer to the corresponding text. It was rather difficult to switch back and forth between text and tables.

**Is the rationale for developing the new method (or application) clearly explained?**
Partly

**Is the description of the method technically sound?**
Yes

**Are sufficient details provided to allow replication of the method development and its use by others?**
Yes

**If any results are presented, are all the source data underlying the results available to ensure full reproducibility?**
Partly

**Are the conclusions about the method and its performance adequately supported by the findings presented in the article?**
Yes

***Competing Interests:*** No competing interests were disclosed.

***Reviewer Expertise:*** My areas of expertise include semantic web technologies as well as geospatial data and query processing.

**I confirm that I have read this submission and believe that I have an appropriate level of expertise to confirm that it is of an acceptable scientific standard, however I have**

**significant reservations, as outlined above.**

**Antonis Troumpoukis**, National Center for Scientific Research (NCSR) Demokritos, Ag. Paraskevi, Greece

Regarding the comment on the (lack of) adoption of geospatial RDF/SW technologies, and on GeoSPARQL losing ground, we would like to react by pointing out that GeoSPARQL is an OGC standard; That goes to show that, although W3C was also naturally involved, the main drive for developing the standard comes from the geospatial databases community. GeoSPARQL has also been implemented by Oracle, with the relevant reference added in the article. In any case, we would also like to point out that the work described here is shown to lead to greater efficiency, thus creating an opportunity for greater adoption of linked geospatial data in general and of federated linked geospatial data in particular. This would allow GIS technologies (including mature systems such as Oracle, with a demonstrated interest in GeoSPARQL) to also participate in the decentralization drive of the modern Web. We have rephrased the conclusions to more clearly spell out that we need to complement the work described here with work on predicting the run-time complexity of queries. This is clearly future work that relates to the query optimizer (rather than the source selector), so we give some initial thoughts in this direction in the final section. We have expanded the related work section with more discussion on relevant literature and how our work is positioned with respect to it. We have also expanded the background, including adding a reference to RCC8. Regarding the relationship between void:Dataset, qb:Dataset, and dcat:Dataset we feel that void:Dataset is the most appropriate term since it aims at annotating a SPARQL endpoint with useful statistics and summaries. qb:Dataset is too restrictive, since it also prescribes internal structure. dcat:Dataset is too generic, as it is not restricted to SPARQL endpoints but can be used for any data that is available for downloading. We feel that adding such minute SW details to this article would be a regression, but we can add it if you insist with your comment. We do not understand why the data underlying the results are commented as being partly available. All data and software needed to reproduce the experiments is published online. Regarding the positioning of the figures and tables, I am afraid that this is done by the journal's editors, outside our control. We have fixed several typos and broken long sentences into shorter, more readable ones.

***Competing Interests:*** No competing interests were disclosed.

? **Ozgu Can**

Department of Computer Engineering, Faculty of Engineering, Ege University, Bornova, Turkey

The manuscript proposes a geospatial source selector for geospatial data, presents the implementation of the related source selector and the correctness of the proposed source selection. Also, evaluation results are compared.

The manuscript should address the following issues before indexing:

The motivation is incomplete. In the Introduction, the problems that the manuscript is focused on should be expressed in more detail with solid examples. Moreover, the state of the art should be discussed in more detail. The unique contribution needs to be clarified. For this purpose, contributions should be listed in a subsection. The main problems of federated query processing should also be presented and discussed. Also, the background section could be improved (such as with federation approaches, etc). Real-life use cases should be given and the proposed study should be discussed by considering the widespread effect of the proposed source selection. The literature work should be elaborated more. Also, recent studies in the field should be added and the review of the related works should be detailed. Further, some illustrations could be added to strengthen the presented manuscript, such as an example of a GeoSPARQL query, the overall architecture of the query processing and source selection, etc. The organization and readability of the manuscript needs to be improved.

**Is the rationale for developing the new method (or application) clearly explained?**
Partly

**Is the description of the method technically sound?**
Partly

**Are sufficient details provided to allow replication of the method development and its use by others?**
Yes

**If any results are presented, are all the source data underlying the results available to ensure full reproducibility?**
Yes

**Are the conclusions about the method and its performance adequately supported by the findings presented in the article?**
Yes

*Competing Interests:* No competing interests were disclosed.

*Reviewer Expertise:* Semantic Web, Knowledge Management, Ontology Engineering, Security, Privacy,

**I confirm that I have read this submission and believe that I have an appropriate level of**

**expertise to confirm that it is of an acceptable scientific standard, however I have significant reservations, as outlined above.**

Author Response 29 Sep 2022

**Antonis Troumpoukis**, National Center for Scientific Research (NCSR) Demokritos, Ag. Paraskevi, Greece

We have added an explicit list of contributions in the "Introduction" section. We have added material that motivates our work in the "Motivation and use case" section. We have expanded the related work section with more discussion on relevant literature and how our work is positioned with respect to it. We have also expanded the background with a more elaborated presentation of federated query processing systems (including architecture diagram). Finally, we have added an illustrative example of a GeoSPARQL query in the main methodological section. Real-life use cases are limited by the lack of technical infrastructure to implement them, infrastructure which is what we are presenting in this article. But, as we note in the article, it is a real-world phenomenon that datasets are partitioned either by jurisdiction or by gridding. This motivates our experimental setup, where we have successfully federated unaligned layers where one layer is partitioned by jurisdiction and another by gridding.

***Competing Interests:*** No competing interests were disclosed.